\documentclass[twoside]{ilcws07}
\usepackage[latin1]{inputenc}
\usepackage[dvips]{graphicx,epsfig,color}
\usepackage{wrapfig,rotating}
\usepackage{amssymb,amsmath,array}

\begin{document}
\title{
Two-Loop Electroweak NLL Corrections: \\
from Massless to Massive Fermions} 
\author{Bernd Jantzen
\thanks{Talk based on work done in collaboration with A.~Denner
  and S.~Pozzorini.}
\vspace{.3cm}\\
Paul Scherrer Institut (PSI) \\
5232 Villigen PSI - Switzerland
}

\maketitle

\rightline{\raisebox{6cm}[0cm]{PSI-PR-07-04}}

\begin{abstract}
Recently the two-loop next-to-leading logarithmic (NLL)
virtual corrections to arbitrary processes with massless external
fermions have been calculated.
Within the spontaneously broken electroweak theory the one- and
two-loop mass singularities have been derived to NLL accuracy and
expressed as universal correction factors depending only on the
quantum numbers of the external particles.
This talk summarizes the results for massless fermionic processes
and presents new aspects arising in the extension of
the corresponding loop calculations to massive external fermions.
As a preliminary result, the Abelian form factor for massive fermions
is given.
\end{abstract}

\section{Electroweak corrections at high energies}

Past and present collider experiments have explored high-energy
processes at energy scales at the order of or below the masses $M_W$
and $M_Z$ of the weak gauge bosons.
But the Large Hadron Collider (LHC) and the proposed International
Linear Collider (ILC) will reach scattering energies in the TeV
regime. For the first time, the characteristic energy~$Q$ of the
reactions will be very large compared to $M_W$.
At these high energies $Q \gg M_W$, electroweak radiative corrections
are enhanced by large logarithms $\ln(Q^2/M_W^2)$, which start to be
sizable at energies of a few hundred GeV and increase with energy.
At LHC and ILC, logarithmic electroweak effects can amount to tens of
per cent at one loop and several per cent at two loops.
In view of the expected experimental precision especially at ILC,
theoretical predictions with an accuracy of about $1\%$ are required,
so the two-loop corrections are crucial.

For sufficiently high~$Q$, mass-suppressed terms of
$\mathcal{O}(M_W^2/Q^2)$ become negligible and the
electroweak corrections assume the form of a tower of logarithms with
terms $\alpha^l \ln^j(Q^2/M_W^2)$, $0 \le j \le 2l$, at $l$ loops.
The leading logarithms (LLs) with power $j=2l$ are known as Sudakov
logarithms~\cite{Sudakov:1954sw}. The subleading logarithms with
$j= 2l-1, 2l-2, \ldots$ are denoted as next-to-leading logarithmic (NLL),
next-to-next-to-leading logarithmic (N$^2$LL) terms, and so on.
The experience with four-fermion
processes~\cite{JKPS:all,Jantzen:2006jv} shows that the subleading
logarithmic contributions may be of the same size as the leading ones.
In addition, large cancellations occur between the individual
logarithmic terms, so the restriction to the LL approximation is not
sufficient, and the NLL corrections or even further subleading terms
are required.

\subsection{Origin of electroweak logarithms}

Logarithms $\ln(Q^2/M_W^2)$ arise from mass singularities, when a
virtual gauge boson (photon~$\gamma$, $Z$ or $W^\pm$ boson) couples to
an on-shell external leg and to any other (internal or external) line
of the diagram.
The region where the gauge boson momentum is collinear to the
momentum of the external particle yields a single-logarithmic
one-loop contribution.
In the special case that the gauge boson is exchanged between two
different external legs, a double-logarithmic contribution arises from
the regions where the gauge boson momentum is soft and collinear to
one of the external momenta.
In addition, ultraviolet (UV) singularities lead to single-logarithmic
contributions.

In the case of photons, the mass singularities are not regulated by a
finite gauge boson mass. In $D = 4-2\epsilon$ space--time dimensions,
the singularities appear as poles $1/\epsilon$ and
$1/\epsilon^2$ per loop.
For a consistent treatment of leading and subleading logarithmic
contributions, each pole in $\epsilon$ has to be counted like a
logarithm $\ln(Q^2/M_W^2)$.
Finite masses of the external particles regularize the collinear
singularities and lead to logarithms involving these masses.

It has been shown at one loop for arbitrary
processes~\cite{Denner:2000jv} and at two loops for massless fermionic
processes~\cite{Denner:2006jr} that the electroweak LL and NLL
corrections are universal: they depend only on the quantum numbers of
the external particles and can be written in terms of universal
correction factors which factorize from the Born matrix element.

\subsection{Approaches for virtual two-loop electroweak corrections at
  high energies}

Two-loop electroweak corrections at high energies have been studied in
recent years with two complementary approaches.
On the one hand, evolution equations known from QCD have
been applied to the electroweak theory by splitting the latter into a
symmetric $SU(2)\times U(1)$ regime above the weak scale~$M_W$ and a
QED regime below the weak scale. Then the evolution equations permit
to resum the one-loop result to all orders in perturbation theory.
From this approach the LL~\cite{Fadin:1999bq} and NLL~\cite{Melles:all}
corrections for arbitrary processes as well as the N$^2$LL
approximation for massless four-fermion processes $f \bar f \to f'
\bar f'$~\cite{KPMS:all} are known, where the NLL and N$^2$LL terms
are valid in the equal mass approximation $M_Z=M_W$.

On the other hand, various calculations have checked and extended the
resummation predictions by explicit diagrammatic two-loop
calculations.
At first, the LLs for the fermionic form factor~\cite{Melles:HKK} were
obtained, then the LLs for arbitrary
processes~\cite{Beenakker:2000kb},
the angular-dependent NLLs for arbitrary
processes~\cite{Denner:2003wi} and
the complete NLLs for the massless fer\-mi\-on\-ic form
factor~\cite{Pozzorini:2004rm}.
Finally, the N$^3$LL approximation for the massless fermionic form
factor was calculated for $M_Z=M_W$ and combined with the evolution
equations, yielding the N$^3$LL corrections for massless
neutral-current four-fermion processes in an expansion $M_Z \approx
M_W$ around the equal mass case~\cite{JKPS:all,Jantzen:2006jv}.

\section{Two-loop next-to-leading logarithmic corrections}

In order to complete the missing diagrammatic NLL calculations,
the goal of this project is to derive virtual two-loop electroweak
corrections for arbitrary processes in NLL accuracy.
In contrast to the resummation approaches, we rely on the complete
spontaneously broken electroweak theory.
We consider processes with external momenta~$p_i$, where all
kinematical invariants, $r_{ij} = (p_i+p_j)^2$, are of
the order of the large scale $Q^2 \gg M_W^2$.
We implement the particle masses $M_W$, $M_Z$, $m_t$ and
$M_{\text{Higgs}}$, which are different, but of the same order. In
particular, we consider a massive top quark and neglect the masses of
the other fermions.
We thus get combinations of large logarithms $L = \ln(Q^2/M_W^2)$ and
poles in~$\epsilon$ from virtual photons.
At $l$ loops, terms $\alpha^l L^n \epsilon^{-j+n}$ are LLs if
$j=2l$, and NLLs if $j=2l-1$ ($n=0,1,\ldots$).
The NLL coefficients involve angular-dependent logarithms,
$\ln\bigl(-r_{ij}/Q^2)$, and logarithms of mass ratios,
$\ln(M_Z^2/M_W^2)$ and $\ln(m_t^2/M_W^2)$.

We have completed the calculation for processes with massless external
fermions~\cite{Denner:2006jr} and are about to extend our results to
massive fermionic processes.

\subsection{Extraction of NLL mass singularities}

In order to extract the mass singularities from the loop diagrams, we
first isolate the so-called factorizable contributions: These are
diagrams where the gauge bosons couple only to external legs, not to
internal legs of the tree subdiagram, and where the gauge boson
momenta have been set to zero in the tree subdiagram.
For these factorizable contributions we use a soft--collinear
approximation which eliminates the Dirac structure of the
loop corrections and factorizes the loop integrals from the
Born matrix element. This approximation is an extension of the
eikonal approximation and reproduces the correct NLL result not only
for soft, but also for collinear gauge bosons.

The remaining non-factorizable contributions are obtained by
subtracting from all diagrams yielding mass singularities the
factorizable contributions.
We have shown that the non-factorizable contributions vanish
due to the collinear Ward identities proven in~\cite{Denner:2000jv}.

\begin{figure}[ht]
  \centering
  \includegraphics{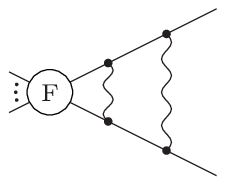} \hspace{-0.4cm}
  \includegraphics{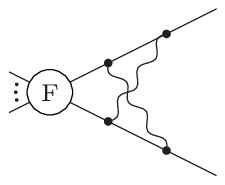} \hspace{-0.4cm}
  \includegraphics{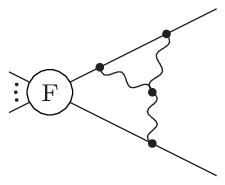} \hspace{-0.4cm}
  \includegraphics{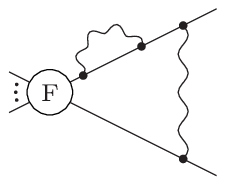} \hspace{-0.4cm}
  \includegraphics{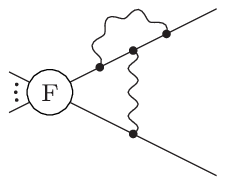}
  \\
  \includegraphics{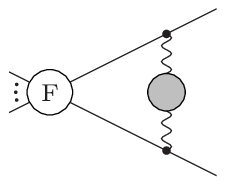} \hspace{-0.4cm}
  \includegraphics{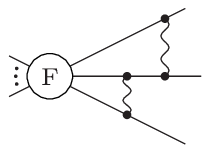} \hspace{-0.4cm}
  \includegraphics{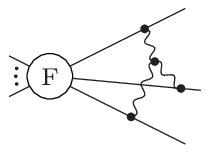} \hspace{-0.4cm}
  \includegraphics{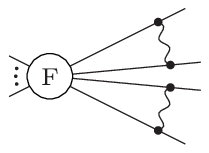}
  \caption{Two-loop factorizable contributions for massless external
    fermions. ``F'' denotes the factorized tree subdiagram, in which
    the gauge boson momenta are set to zero.
    The grey blob in the gauge boson propagator stands for all
    possible self-energy insertions.}
  \label{Fig:2loopfact}
\end{figure}

Therefore only the factorizable contributions need to be evaluated
explicitly. For the LL and NLL terms at two loops, we need a
double-logarithmic contribution from a soft and collinear gauge boson
which is exchanged between two different external legs,
and another, at least single-logarithmic, loop correction.
The two-loop factorizable contributions in the case of massless
external fermions are depicted in Figure~\ref{Fig:2loopfact}.

The factorizable diagrams also include NLL contributions from
UV momentum regions.
When a subdiagram with a small characteristic scale of the order
$M_W^2$ yields UV singularities which are renormalized at the
scale~$Q^2$, large logarithms $\ln(Q^2/M_W^2)$ arise.
The soft--collinear approximation mentioned above is
not valid for UV momenta, so we cannot use it for subdiagrams of this
type and employ projection techniques instead.

\subsection{Results for massless fermionic processes}

We have evaluated the loop integrals of the factorizable contributions
with two independent methods:
An automatized algorithm which is based on the sector decomposition
technique~\cite{Denner:2004iz},
and the method of expansion by regions combined with Mellin--Barnes
representations (see~\cite{Jantzen:2006jv} and references therein).
The NLL result for massless fermionic processes $f_1 f_2 \to f_3 \cdots
f_n$ has been published in~\cite{Denner:2006jr}. It allows to write the
combined one- and two-loop result in the factorized form
$\mathcal{M} = \mathcal{M}_0 F^{\text{sew}} F^Z F^{\text{em}}$,
where $\mathcal{M}_0$ is the Born matrix element,
and the correction terms read
$F^{\text{sew}} = \exp\bigl[\tfrac{\alpha}{4\pi} F_1^{\text{sew}}
  + \bigl(\tfrac{\alpha}{4\pi}\bigr)^2 G_2^{\text{sew}}\bigr]$,
$F^Z = 1 + \tfrac{\alpha}{4\pi} \Delta F_1^Z$
and
$F^{\text{em}} = \exp\bigl[\tfrac{\alpha}{4\pi} \Delta F_1^{\text{em}}
  + \bigl(\tfrac{\alpha}{4\pi}\bigr)^2 \Delta G_2^{\text{em}}\bigr]$.
The symmetric-electroweak factor $F^{\text{sew}}$ equals the result
from a symmetric $SU(2)\times U(1)$ theory where all gauge boson
masses are equal to~$M_W$.
The factor $F^Z$ incorporates the terms from the mass difference $M_Z
\ne M_W$.
And the electromagnetic terms in $F^{\text{em}}$ factorize and
exponentiate separately, such that a separation of the singularities
due to the massless photon is possible.
The one-loop terms $F_1^{\text{sew}}$ and $\Delta F_1^{\text{em}}$ get
exponentiated, and the additional two-loop terms $G_2^{\text{sew}}$
and $\Delta G_2^{\text{em}}$ are proportional to $\beta$-function
coefficients.
For details of the correction terms, we refer to~\cite{Denner:2006jr}.

Our results confirm the resummation predictions based on
the evolution equations.
By applying our general correction factors to the case of massless
four-fermion scattering, we have found agreement with the
neutral-current results in~\cite{JKPS:all,KPMS:all}, and we have
obtained a new NLL result for the charged-current processes.

\section{From massless to massive fermions}

For massive external fermions, the diagrams from the
factorizable contributions have to be reevaluated,
additional diagrams with Yukawa interactions have to be considered
and the cancellation of the non-factorizable contributions must be
verified.
This section deals with new complications which arise from massive
external fermions in the loop integrals.

\subsection{Expansion by regions with massive external particles}

Expansion by regions~\cite{Smirnov:EbR,Smirnov:2002pj}
is a powerful method for the asymptotic expansion
of loop integrals. It is based on the following recipe:
Divide the integration domain of the loop momenta into regions
corresponding to the asymptotic limit considered.
In every region, expand the integrand appropriately.
Integrate each of the expanded terms over the whole integration
domain.

The integrand is expanded before integration, and each
expanded term has a unique order in powers of the large scale $Q$
and the small scale $M_W$.
But on-shell momenta~$p_i$ of massive external particles involve two scales,
as their momentum squared is $p_i^2=m_i^2\sim M_W^2$ and their
combinations with other external momenta are $r_{ij} = (p_i+p_j)^2
\sim Q^2$.
In order to separate these two scales, the external momenta are
reparametrized in terms of light-like momenta~$\tilde p_i$ as
$p_i = \tilde p_i + (p_i^2/\tilde r_{ij}) \tilde p_j$,
with some other external leg $j\ne i$ and $\tilde p_i^2=\tilde p_j^2=0$,
$\tilde r_{ij} = 2\tilde p_i \tilde p_j$~\cite{Smirnov:2002pj}.
Through this shift, all contractions of external momenta with loop
momenta can now be divided into parts of distinct scales, and the
expansion is done in inverse powers of the new large scales
$\tilde r_{ij} = r_{ij} + \mathcal{O}(M_W^2)$.

With respect to any pair of external light-like momenta $\tilde p_i,
\tilde p_j$, the loop momenta can be expressed in Sudakov components
parallel and perpendicular to these external momenta:
$k = k^{(i,j)}_i \tilde p_i/Q + k^{(i,j)}_j \tilde p_j/Q
  + k^{(i,j)}_\perp$,
with $k^{(i,j)}_i = 2\tilde p_j k\,Q/\tilde r_{ij}$,
$k^{(i,j)}_j = 2\tilde p_i k\,Q/\tilde r_{ij}$ and
$\tilde p_i k^{(i,j)}_\perp = \tilde p_j k^{(i,j)}_\perp = 0$.
In each region, the components of the loop momenta are
assigned specific sizes in powers of $Q$ and $M_W$.
Typical regions are listed in Table~\ref{Tab:regions}.
\begin{table}[ht]
  \centering
  \begin{tabular}{|c||c|c|c|c|c|c|c|}
  \hline
    Region &
      hard & soft & $i$-collinear & $j$-collinear &
      ultrasoft & $i$-ultracoll. & $j$-ultracoll. \\
  \hline
    $k^{(i,j)}_i$ &
      $Q$ & $M_W$ & $Q$ & $M_W^2/Q$ &
      $M_W^2/Q$ & $M_W^2/Q$ & $M_W^4/Q^3$ \\
  \hline
    $k^{(i,j)}_j$ &
    $Q$ & $M_W$ & $M_W^2/Q$ & $Q$ &
    $M_W^2/Q$ & $M_W^4/Q^3$ & $M_W^2/Q$ \\
  \hline
    $k^{(i,j)}_\perp$ &
    $Q$ & $M_W$ & $M_W$ & $M_W$ &
    $M_W^2/Q$ & $M_W^3/Q^2$ & $M_W^3/Q^2$ \\
  \hline
  \end{tabular}
  \caption{Typical regions with the corresponding sizes of loop
    momentum components}
  \label{Tab:regions}
\end{table}%
While the hard, soft, collinear and ultrasoft regions are already
present for massless external particles, the two ultracollinear
regions are only relevant for massive external particles.

\subsection{Power singularities and fermion masses}

Asymptotic expansions with small masses and large kinematical scales
not only produce logarithmic mass singularities, but also power
singularities $Q^2/M_{W,Z}^2$ and $Q^2/m_t^2$.
These are generated at two loops by
subdiagrams with a small scale of the order $M_W^2$.
The method of expansion by regions predicts, for the contribution of
each region, where power singularities can appear, by means of a
simple power counting in the expanded integrals.

When complete Feynman diagrams are considered, the terms in the numerator
ensure the cancellation of the power singularities.
In diagrams where power singularities are present for
individual scalar integrals, care must be taken to keep all the mass
factors in the numerator which ensure the cancellations.
In particular, the masses in the numerator of fermion propagators and
in the Dirac equation of the spinors may not be neglected.
Therefore we are not allowed to use the soft--collinear approximation
for small-scale subdiagrams. However, these are exactly the same
diagrams where we have employed alternative projection techniques
already in the massless case in order to get the UV contributions
right.

Additional complications originate from fermion masses in the
numerator due to the chiral structure of the electroweak theory. With
each mass factor along a fermion line, the chirality of the fermion in
its interactions with the weak gauge bosons changes. We have found,
though, that fermion masses in the numerator are relevant exclusively
in pure QED diagrams where the chirality changes do not matter.

\subsection{Preliminary results}

We have completed the calculation of all factorizable contributions
involving two massive or massless external fermion legs. This permits
to determine the two-loop form factor in an Abelian model
with both a massive gauge boson (mass~$M_W$, coupling~$\alpha$) and a
massless one (coupling~$\alpha'$).
The one-loop form factor as a function of the two external fermion
masses is given by
$F_1(m_1,m_2) = \frac{\alpha}{4\pi} F_1^M
  + \frac{\alpha'}{4\pi} \bigl[ F_1^0(0,0) + \Delta F_1^0(m_1)
  + \Delta F_1^0(m_2) \bigr]$.
The NLL contribution (up to the order~$\epsilon^2$) from the massive
gauge boson is independent of the fermion masses,
\begin{equation}
  F_1^M =
  -L^2 - \frac{2}{3} L^3 \epsilon - \frac{1}{4} L^4 \epsilon^2
  + 3 L + \frac{3}{2} L^2 \epsilon + \frac{1}{2} L^3 \epsilon^2,
\end{equation}
with $L = \ln(Q^2/M_W^2)$,
while the contribution from the massless gauge boson is split into a
completely massless part and corrections for each of the fermion
masses:
\begin{align}
  F_1^0(0,0) &= -2\epsilon^{-2} - 3\epsilon^{-1}, \quad
  \Delta F_1^0(0) = 0,
\nonumber \\*
  \Delta F_1^0(m_i) &=
  \epsilon^{-2} + L_i \epsilon^{-1} + \frac{1}{2} L_i^2
  + \frac{1}{6} L_i^3 \epsilon + \frac{1}{24} L_i^4 \epsilon^2
  + \frac{1}{2} \epsilon^{-1} + \frac{1}{2} L_i
  + \frac{1}{4} L_i^2 \epsilon + \frac{1}{12} L_i^3 \epsilon^2,
\end{align}
with $L_i = \ln(Q^2/m_i^2)$.
We have found that the NLL two-loop form factor (without closed fermion
loops) simply exponentiates the one-loop result,
$F_2(m_1,m_2) = \frac{1}{2} \bigl[ F_1(m_1,m_2) \bigr]^2$.

\section{Conclusions}

We evaluate two-loop electroweak corrections in NLL accuracy for
arbitrary processes with massive and massless external fermions.
The methods which we have successfully applied for massless fermions
work well also in the massive case, and the complications arising from
fermion masses are under control. Preliminary results are already
available for the form factor, they factorize and exponentiate like in
the massless case.
The calculation for processes with external fermions will soon be
completed, and our method can be extended to arbitrary processes
involving external gauge bosons or scalar particles.

\section*{Acknowledgments}

The author gratefully acknowledges the pleasant collaboration with
A.~Denner and S.~Pozzorini in the project presented in this talk.


\begin{footnotesize}

\end{footnotesize}


\end{document}